# Multi-frequency MRE for elasticity quantitation and optimal tissue discrimination: a two-platform liver fibrosis mimicking phantom study

Andoh F.[1], Yue J.L.[1], Julea F.[2], Tardieu M.[3], Noûs C.[4], Page G.[2], Garteiser P.[2], Van Beers B.E.[2,5], Maître X.[1], and Pellot-Barakat C.[1]

[1] Université Paris-Saclay, CEA, CNRS, Inserm, BioMaps, Orsay, France
[2] Laboratory of Imaging Biomarkers, Center for Research on Inflammation, UMR 1149, Inserm, Université de Paris, Paris, France
[3] Montpellier Cancer Research Institute (IRCM), Inserm, University of Montpellier, Montpellier, France
[4] Laboratoire Cogitamus
[5] Department of Radiology, Beaujon University Hospital Paris Nord, AP-HP, Clichy, France

E-mail: claire.barakat@inserm.fr

## Abstract

In the framework of algebraic inversion, Magnetic Resonance Elastography (MRE) repeatability, reproducibility and robustness were evaluated on extracted shear velocities (or elastic moduli). The same excitation system was implemented at two sites equipped with clinical MR scanners of 1.5 T and 3 T. A set of four elastic, isotropic, homogeneous calibrated phantoms of distinct elasticity representing the spectrum of liver fibrosis severity was mechanically characterized. The repeatability of the measurements and the reproducibility between the two platforms were found to be excellent with mean coefficients of variations of 1.62 % for the shear velocity mean values and 1.95 % for the associated standard deviations. MRE velocities were robust to the amplitude and pattern variations of the displacement field with virtually no difference between outcomes from both magnets at identical excitation frequencies even when the displacement field amplitude was 6 times smaller. However, MRE outcomes were very sensitive to the number of voxels per wavelength, $s$, of the recorded displacement field, with relative biases reaching 62 % and precision losing up to a factor 23.5. For both magnetic field strengths, MRE accuracy and precision were largely degraded outside of established conditions of validity ($6 \lesssim s \lesssim 9$) resulting in estimated shear velocity values not significantly different between phantoms of increasing elasticity. When fulfilling the spatial sampling conditions, either prospectively in the acquisition or retrospectively before the reconstruction, MRE produced quantitative measurements that allowed to unambiguously discriminate, with infinitesimal p-values, between the phantoms mimicking increasing severity of liver fibrosis.

**List of abbreviations:**
AIDE: Algebraic Inversion of the Differential Equation
SWE: Shear Wave Elastography
ICC: Intraclass Correlation Coefficient
CV: Coefficient of Variation
MAPE: Minimum Absolute Percentage Error

The data that support the findings of this study are available in the supplementary material of this article.

## 1. Introduction

Magnetic Resonance Elastography (MRE) aims at mapping the mechanical properties of biological tissues by recording the displacement fields generated by a mechanical wave travelling through them. The extracted viscoelasticity moduli can be advantageously used in clinical diagnosis as the development of most pathological processes comes with an alteration of the tissue mechanical properties.[1–4] In spite of successful applications in the clinic, the spread of MRE is undermined by the lack of accuracy and precision of the measurement on a voxel basis.

MRE outcomes are conditioned by the mechanical waves travelling throughout the targeted tissue (their frequency, amplitude, and pattern within the definite boundary conditions), the acquisition parameters (the voxel size, the motion sensitizing gradients, and the resulting signal-to-noise ratio for a given MR pulse sequence), and the reconstruction method. In the framework of algebraic inversion of the differential equation of motion, AIDE,[5,6] once temporal sampling is set, the factors determining the accuracy and precision of MRE measurements can ultimately be subsumed with two parameters that essentially characterize how well the propagating shear wave is sampled: the spatial sampling factor (or number of voxels per wavelength), $s = \lambda/a$, and the amplitude sampling factor (or data quality factor), $Q = q/\Delta q$, where $\lambda$ is the shear wavelength, $a$, the voxel size, $q$, the amplitude of the curl of the displacement field, and $\Delta q$, the associated measurement uncertainty.[7] These two dimensionless factors can be extracted for any voxel. If $s$ is too small (generally smaller than 6) then the shear wave pattern is undersampled and the shear velocity and elasticity are overestimated. If $s$ is too big (generally greater than 9), then the shear wave pattern is oversampled and the shear velocity and elasticity are underestimated. Otherwise, when it stands in between, the shear wave pattern is optimally sampled and the measurement uncertainty is minimized.[7–9] The amplitude sampling factor, $Q$, increases with $q$ and SNR (along with $1/\Delta q$). For a given shear wave amplitude, the higher the SNR, the smaller the measurement uncertainty is and the less influence $s$ has on this uncertainty. Conversely, the lower the SNR, the higher the measurement uncertainty and the more sensitive to $s$ the measurement precision and accuracy are. With a rather low SNR of 8, simulations predicted and experiments recorded relative velocity biases of 45 % and threefold standard deviations when the voxel size, all things being equal otherwise, was either halved or doubled with respect to the optimal wave spatial sampling for MRE at excitation frequency $f_{exc} = 2$ kHz, respectively leading to $s \simeq 12$ and $s \simeq 3$.[10] Special attention should therefore be paid when performing MRE either using different voxel sizes and different excitation frequencies,[11,12] simply probing tissues at different stages of a disease,[13,14] or mapping mechanically heterogeneous tissues.[4,15] All those situations exhibit different wavelengths, leading to different spatial sampling conditions with different measurement bias and precision. Furthermore, additional mechanical features are expected to be revealed by the biological tissue dispersive behavior.[16,17] In this framework, MRE rheological studies carried out by merely sweeping the excitation frequency present dispersion laws that are inherently flawed by frequency-dependent measurement bias and precision.

In this work, we aimed at investigating the repeatability, reproducibility, robustness, accuracy and precision of MRE along optimal sampling conditions across two MRI platforms at 1.5 T and 3 T in two different sites. For that purpose, multi-frequency experiments were carried out on mechanically-calibrated phantoms that mimic liver fibrosis. The data were acquired under controlled conditions with the same excitation device and MR pulse sequence parameters at both sites. All acquired data were processed with the same reconstruction method. Estimated MRE shear velocities were compared to ultrasound shear wave elastography (SWE) measurements, a competing imaging modality in the liver, for accuracy evaluation. The ability to discriminate between shear velocity estimates in optimal and non-optimal conditions was evaluated. A re-conditioning strategy was considered to cope with non-optimally acquired datasets and resulting shear velocity estimates were compared with values obtained with the optimally pre-conditioned datasets.

## 2. Experimental

### 2.1. Test phantoms

The test phantoms (C1, C2, C3, C4) consisted of four cylinders housing 10 cm diameter, 12 cm height, homogeneous Zerdine® solid hydrogel (Model 039, CIRS, Arlington, VA, USA). These phantoms were developed and validated in a study sponsored by the Quantitative Imaging Biomarker Alliance. They served as reference standards to evaluate shear wave velocity measurements with quasi-static compression dynamic mechanical analysis, vibration-controlled transient elastography, and hyper-frequency viscoelastic spectroscopy. They were shown to be more elastic than viscous.[18]

The four phantoms (C1, C2, C3, C4), as characterized by the manufacturer, provided respective Young's moduli $E$ of 3.5, 11.4, 28.6, and 44.8 kPa that matched those of liver fibrosis from normal state to cirrhosis (Figure 1). With a density of 1030 kg · m$^{-3}$ and a Poisson's ratio of 0.5, expected shear elasticity moduli are roughly 1.7, 3.8, 9.5, and 14.9 kPa and shear velocities, 1.1, 1.9, 3.0, and 3.8 m · s$^{-1}$.

The phantoms were also mechanically characterized by SWE with an Aixplorer ultrasound system (Supersonic Imagine, Aix-en-Provence, France) using three different ultrasonic probes (XC6-1, SL10-2, and SL15-4). Voxels were anisotropic with 2D resolutions (axial × lateral sizes) of (0.8 × 0.6) mm$^2$, (0.35 × 0.20) mm$^2$ and (0.21 × 0.20) mm$^2$ for XC6-1, SL10-2, and SL15-4 respectively. Elevation resolutions at the focal point where the beam was the thinnest were 1.62 mm, 1.16 mm, 1.05 mm for XC6-1, SL10-2, and SL15-4. During acquisition, an artificial arm steadily held the ultrasound probes on the surface of the phantoms to minimize operator influence. The insonification window was placed in the center of the phantoms and its size was adapted to the frequency bandwidth of each probe. SWE velocity maps were processed together with the corresponding quality maps provided by the Aixplorer. Quality map values ranged from 0 to 1, 1 expressing a perfect confidence in the estimated shear velocity. Regions with quality below 0.7 were masked out before calculating the mean and standard deviation values of the shear velocity for the three ultrasound probes.[19]

### 2.2. MRE acquisition

MRE bench setup and sequence

MRE acquisitions were carried out on Achieva 1.5 T and Ingenia 3 T MR systems (Philips Healthcare, Eindhoven, The Netherlands). The phantoms were placed at the center of the magnet bore into an 8 channel SENSE knee coil (Philips Healthcare, Eindhoven, The Netherlands) with their axis horizontally aligned with the directing magnetic field (Figure 2).

A point source mechanical excitation was induced in the phantom by guided pressure waves through a 1 mm air-tight diameter acoustic adapter (Figure 2). The pressure waves were remotely generated from the technical room with a function generator (AFG 3021B, Tektronix, Beaverton, OR, USA) before being amplified with a power amplifier (P5000S, Yamaha, Shizuoka, Japan), transduced with a 300 W 12" woofer (PHL Audio 4530, Chartrettes, France), and guided through the Faraday cage along 22 mm inner diameter, 6.24 m long altuglas® tubes (Altuglas, La Garenne-Colombes, France) and adapting hoses to the surface of the phantom (Figure 2).[20] The generation of pressure waves was trigged by the MRI system for synchronization with the MRE acquisition and monitored with an oscilloscope (TDS-2014, Tektronix, Beaverton, OR, USA).

The source pressure generated at the surface of the phantom was monitored from the MRI console room and recorded onsite with an optical fiber sensor at 5 kHz (EVO-RM-8, FISO, Québec, QC, Canada). The whole MRE bench was designed to facilitate reproducibility of the point source amplitudes and patterns of the mechanical waves induced throughout all phantoms and MRI systems.

Amplitudes of the pressure waves were set for each frequency to provide easily measurable waves within the phantoms while avoiding as much as possible reflections on the cylindrical walls. The wave measurability was assessed on preliminary acquisitions by visual inspection of the MR phase images. Limiting wave interferences in the phantom was important to avoid too much amplitude difference between potential node and antinode regions. Selected pressure amplitudes recorded at the surface of the phantoms were repeated and reproduced for each experiment performed at the same frequency. For the two MRI platforms, the acoustic resonances of the closed wave-guiding system were characterized by wobulation from 10 Hz to 400 Hz.

A standard multi-slice motion-encoding spin-echo sequence was implemented with a field of view FOV = (120 × 120 × 30) mm$^3$ covering the upper part of the cylindrical phantoms, a matrix size of (96 × 96 × 24), and an isotropic voxel of size $a$ = 1.25 mm. The amplitudes of the motion-encoding gradients set by the MR systems were very close (21 mT · m$^{-1}$ at 1.5 T and 22.5 mT · m$^{-1}$ at 3 T). Hence the motion sensitivity could be considered similar for the two MR systems. The number of bipolar motion-encoding gradients, $N_{MEG}$, was increased up to sixfold when the frequency of the mechanical vibration varied from 40 Hz to 320 Hz so as to compensate, at least partly, the higher wave attenuation at higher frequencies (Table 1). $TE$ and $TR$ values were imposed by the frequency and number of motion-encoding gradients. $TE$ ranged from 27 and 62 ms while the repetition time $TR$ concomitantly varied between 1,049 and 1,800 ms in order to maintain a relatively consistent SNR (Table 1).

MRE Experiments

The phantom C2 (11.4 kPa) was first mechanically characterized with both MR systems at seven different excitation frequencies selected close to the system resonances ($f_{\text{exc}} = \{40, 60, 90, 130, 175, 207, 320\}$ Hz). The acquisition at 130 Hz that empirically presented the best wave patterns, was repeated at the end of the experimental runs for repeatability assessment.

The three other phantoms were then mechanically characterized with both MR systems at an excitation frequency close to their corresponding expected optimal excitation frequency, $f_{\text{opt}}$, for a voxel size $a$ of 1.25 mm. It was derived from preliminary multi-frequency MRE acquisitions.[19] The reference values for the shear velocity of C1, C3 and C4 were $0.955 \text{ m} \cdot \text{s}^{-1}$, $2.141 \text{ m} \cdot \text{s}^{-1}$ and $3.320 \text{ m} \cdot \text{s}^{-1}$, which, given $s \in [6, 9]$, led to optimal frequency ranges of $[85, 127]$ Hz, $[190, 285]$ Hz and $[295, 443]$ Hz for C1, C3 and C4 respectively. Acoustic resonance frequencies of the excitation system were chosen close to these ranges for each phantom: $f_{\text{opt}} = \{60, 207, 320\}$ Hz for C1, C3 and C4 respectively. As most liver MRE studies in the literature report excitation frequencies of 60 Hz, data were also acquired at $f_{\text{conv}} = 60$ Hz and 1.5 T for every phantom.

Moreover, all the complex MR raw data acquired in non-optimal conditions were resampled before phase unwrapping to retrospectively achieve optimal $s$ conditions before extraction of the displacement fields and computation of the mechanical properties. Up and downsampling were performed through a Lanczos kernel. The resampling factors were given by the ratio $f_{\text{exc}}/f_{\text{opt}}$ between the excitation frequency, $f_{\text{exc}}$, and the optimal frequencies, $f_{\text{opt}}$ – for C2 between 40 and 320 Hz and C1-C4 at 60 Hz – leading to an optimal pre-reconstruction voxel size $a_{\text{opt}} = a \cdot f_{\text{exc}}/f_{\text{opt}}$. Lanczos kernel widths were adapted to each downsampling factor to match expected SNR gain one would obtain with Gaussian noise by mere averaging over downsampled voxels.

### 2.3. MRE reconstruction

The components of the 3D displacement field $u_i(\mathbf{r}, t)$, with $i = \{x, y, z\}$, of a voxel located at $\mathbf{r}$ and taken at time $t$ can be computed from the recorded MRI phase values $\varphi_i(\mathbf{r}, t)$:

$$\varphi_i(\mathbf{r}, t) = \gamma \frac{N}{2} T \cdot A_{MEG,i} \cdot u_i(\mathbf{r}, t) \quad (1)$$

with $\gamma$, the gyromagnetic ratio of hydrogen nuclei, $N$, the number of bipolar motion-encoding gradient of duration $T = 1/f$, and $A_{MEG,i}$, the amplitude of the motion-encoding gradient along the three directions $i$.

The components $q_i(\mathbf{r}, t)$ of the curl of $\mathbf{u}(\mathbf{r}, t)$ satisfy the Helmholtz equation (2) for shear wave in a locally homogeneous isotropic viscoelastic medium with complex shear modulus $G^*$:

$$\rho \omega^2 q_i(\mathbf{r}, t) = G^* \nabla^2 q_i(\mathbf{r}, t)$$

where $\omega = 2\pi f$, $\mathbf{q}(\mathbf{r}, t) = \nabla \times \mathbf{u}(\mathbf{r}, t)$ and $i = \{x, y, z\}$

By algebraic inversion, the shear dynamic, $G_i'$, and loss, $G_i''$, moduli can be deduced along each spatial dimension $i$. As the phantoms are mainly elastic, we will here focus on $G_i'$ as well as on the shear wave velocity, $V_{s,i} = \sqrt{G_i'/\rho}$. For each voxel, the average velocity over the three directions was weighted by the data quality factor $Q$ according to:

$$V_s = \frac{1}{Q^2} \sum_i \left( \left( \frac{q_i}{\Delta q_i} \right)^2 V_{s,i} \right)$$

$$Q = \sum_i \left( \frac{q_i}{\Delta q_i} \right)^2$$

with $\Delta q_i$ being the uncertainty on the amplitude $q_i$ of the curl component of the displacement field $q_i(\mathbf{r}, t)$. The $Q$-weighted average velocity $V_s$ defined in equation (3) minimizes the minimum absolute percentage error (MAPE) and thus provides the most accurate and precise estimation of the local shear velocity with AIDE.[7]

Masks were automatically generated by magnitude thresholding at a tenth of the maximal signal for the different acquisitions. Data were eroded to exclude possible reconstruction biases at the boundaries. The size of the erosion kernel was chosen by checking when the standard deviation of the shear velocity became stable. For $a = 1.25$ mm, the latter reached a plateau when the velocity map had been eroded by 6 voxels. After erosion, the intersection of individual phantom masks was computed to provide a unique mask for further analysis. Both the mean value and the standard deviation of the amplitudes of the generated displacement fields, $\langle A \rangle$ and $\sigma_A$, were considered at every frequency for a fair evaluation of the inhomogeneous amplitude of the spherical wave throughout each phantom. Similarly, mean values and standard deviations of $Q$ were considered to reflect this inherent inhomogeneity. Mean values, $\langle \text{SNR} \rangle$, $\langle V_s \rangle$, $\langle G' \rangle$ and associated standard deviations were calculated over the resulting phantom mask to exhibit the accuracy and the precision of the MRE acquisition and reconstruction. For the sake of simplicity, we will mainly refer from now on to the shear velocities but the corresponding shear dynamic moduli could have been used instead.

In order to compare MRE outcomes to the manufacturer's calibration values as well as to the SWE values, the Young's modulus $E$ was deduced for each phantom using the simple relationship for homogeneous, isotropic, elastic, and incompressible media:

$$E = 3G' \tag{5}$$

### 2.4. Data analyses

Statistical analyses were performed using the mean shear velocity, $\langle V_s \rangle$, as primary outcome. Boxplots of $V_s$ were computed to visually represent the variability of the data, both within and between experiments conducted with differing acquisition and reconstruction conditions. Non-parametric unpaired two-samples Wilcoxon tests were performed using R.[21] A result was deemed statistically significant if the probability was lower than 1 % or p-value < 0.01.

The measurement repeatability was evaluated based on the voxel-wise shear velocity estimated from the two experiments at 1.5 T and 130 Hz. The inter-platform reproducibility was evaluated by comparing the voxel-wise $V_s$ estimated from identical experiments at 1.5 T and 3 T (same phantom, same excitation frequency). Measurement repeatability and reproducibility were quantified with two-way random effects single-measure intra-class correlation coefficients (ICCs) and coefficients of variation (CVs). These coefficients were computed for the shear velocity mean values, $\langle V_s \rangle$, and standard deviation values, $\sigma_{V_s}$, to assess both accuracy and precision of the measurements.

## 3. Results

### 3.1. Measurement agreement

Measurement repeatability

Repeated measurements at 130 Hz on C2 yielded identical shear velocities at 1.5 T ($(1.48 \pm 0.04)$ m·s$^{-1}$) and very similar values at 3 T ($(1.49 \pm 0.05)$ m·s$^{-1}$ and $(1.50 \pm 0.05)$ m·s$^{-1}$) (Table 1). With a relative variability of 3 %, velocity maps at 130 Hz were fairly homogeneous.

Acquisition reproducibility

Mean SNRs were expectedly higher at higher field. They ranged from 22 to 30 at 1.5 T and 34 to 45 at 3 T, all phantoms and frequencies considered (Table 1 and Table 2). SNRs were also slightly higher for longer $TRs$ and shorter $TEs$. The SNR maps were rather homogeneous for any excitation frequency with similar grain texture at both fields (Figure 3 and Figure 5). Total amplitudes of the displacement fields and quality factors were much higher at 1.5 T than at 3 T for all frequencies except at 320 Hz. Mean values $\langle A \rangle$ ranged on average from 1.63 to 15.03 µm at 1.5 T and 0.46 to 4.5 µm at 3 T but they could reach 40.6 µm at 90 Hz and 1.5 T. Mean values $\langle Q \rangle$ ranged from 5 to 274 at 1.5 T and from 9 to 63 at 3 T. The $Q$ maps exhibited patterns analogous to those of the $q$ maps.

Reconstruction reproducibility

While SNR, $A$, $q$, and $Q$ mean values and maps differed between 1.5 T and 3 T, shear velocity values remained similar between the two field strengths. For C2 at varying frequencies, recorded mean values stayed within less than 1 % except at the highest frequency (320 Hz) with $\langle V_s \rangle$ of $(2.53 \pm 0.50)$ m·s$^{-1}$ at 1.5 T versus $(2.13 \pm 0.40)$ m·s$^{-1}$ at 3 T (Table 1). Velocity maps also followed the same trend at both fields: they were fairly homogeneous between 90 and 207 Hz whereas they were degraded either at lower or higher frequencies (Figure 3). The same trend was observed at both fields in the voxel-wise velocity as a function of $Q$ with a distribution that similarly evolved with $f$: It narrowed between 90 and 207 Hz whereas it changed shapes and spread at lower and higher frequencies (Figure S1). For C1-C4 at optimal frequencies, $\langle V_s \rangle$ measurements were also positively reproduced between the two field strengths (Table 2). A slight discrepancy occurred at the highest frequencies (207 Hz for C3 and 320 Hz for C4) but this variation remained well below the maximal measurement standard deviation of 8.3 %. The agreement of the mean values and the standard deviations was excellent for C1 and C2.

The overall mean CVs between 1.5 T and 3 T of $\langle V_s \rangle$ and $\sigma_{V_s}$, all phantoms and frequencies combined, were 1.62 % (0-12.1 %) and 1.95 % (0-10.3 %) respectively. The ICCs were 0.98 (95 % confidence interval $0.943 < ICC < 0.995$) and 0.89 (95 % confidence interval $0.660 < ICC < 0.966$) for $\langle V_s \rangle$ and $\sigma_{V_s}$ respectively.

### 3.2. Two-site measurements of the same phantom at varying frequencies

For both field strengths, the shear velocity variability for C2 was the lowest at 175 Hz (variations of 1.3 % at 1.5 T and 2.6 % at 3 T), as clearly depicted on Figure 3. It set the optimal conditions, $s = 6.9$, for the phantom C2 from which $\langle V_s \rangle = (1.51 \pm 0.02)$ m·s$^{-1}$, can be extracted with the best confidence.

MRE shear velocity measurements of C2 as a function of $f_{exc}$ are represented as boxplots on Figure 4 for 1.5 T

(a) and 3 T (b). For both fields, the estimated shear velocity increases as $f_{exc}$ increases and the variability increases as $f_{exc}$ moves further away from $f_{opt}$ = 175 Hz.

Measurement conditioning

When data sets were retrospectively resampled to approach optimal spatial sampling, $a_{opt}$, for C2 at frequencies ranging from 40 Hz to 320 Hz (Figure 4 (c)), the estimated mean shear velocities fell into a much narrower range with, at 1.5 T, a CV between frequencies of 24.3 % with $a$ = 1.25 mm reduced to 4.8 % with $a$ = $a_{opt}$ even though the estimated mean shear velocities were slightly underestimated at 40, 60 and 90 Hz (Table 4). Concurrently, the measurement precisions were improved. The effects are negligible when data were already in the optimal $s$ domain, at 130 and 207 Hz, but they are radical when data were away from it, at 40, 60 and 320 Hz, with revealed relative biases between 3 % and 62 % and precision gains between 4 and 23.5. Exemplarily at 320 Hz, the estimated $\langle V_s \rangle$ was $(2.53 \pm 0.50)$ m · s$^{-1}$ with $a$ = 1.25 mm and $(1.59 \pm 0.12)$ m · s$^{-1}$ with $a_{opt}$ = 0.68 mm (Table 4).

### 3.3. Two-site measurements of all phantoms at conventional and optimal frequencies

Figure 5 shows the SNR, $Q$, $V_s$ and $G'$ maps for the four phantoms (C1-C4) at 1.5 T and 60 Hz, 1.5 T and $f_{opt}$ (60, 175, 207 and 320 Hz for C1, C2, C3 and C4 respectively) and at 3 T and $f_{opt}$.

Conventional 60 Hz frequency

At the conventional frequency $f_{conv}$ = 60 Hz, the shear velocity and elasticity maps were largely inhomogeneous except for C1 for which 60 Hz lies in the optimal frequency range (Figure 5). MRE measurements at $f_{conv}$ for each phantom are represented as boxplots on Figure 6 (c) for 1.5 T. The corresponding mean values and standard deviations are listed in Table 2. Measurements at $f_{conv}$ failed to discriminate between C3 and C4 with p ≃ 0.14 (rank 212,150). Although the Wilcoxon tests revealed significant differences between the other phantoms, they showed high ranks (100 for C1-C2, 117,990 for C2-C3, and 142,430 for C2-C4).

Optimal frequency

For each phantom, the maps were homogeneous when the respective expected optimal conditions as set by the excitation frequency $f_{opt}$ were respected (at both 1.5 T and 3 T). MRE shear velocity measurements in the optimal conditions for each phantom are represented as boxplots on Figure 6 for 1.5 T (a) and 3 T (b). It can be observed that each measurement at $f_{opt}$ provides a clear discrimination on a voxel-basis of the four phantoms. The statistical differences were highly significant (infinitesimal p-values and low ranks of 268 for C2-C3, 2780 for C3-C4, and 0 otherwise).

Stiffness assessment

Velocity measurements performed at $f_{conv}$ largely differed from the measurements at $f_{opt}$. The mean relative difference was only 4.6 % for C2 but went up to 21 % for C3 and to 43 % for C4. The measurement uncertainty was also three to sixfold higher away from the optimal condition at $f_{conv}$ than at $f_{opt}$. Nonetheless, in any configuration, the mean shear velocity increased with the phantom grade. However, at $f_{conv}$, it only increased from 0.83 m · s$^{-1}$ for C1 to 1.88 m · s$^{-1}$ for C4 whereas, at $f_{opt}$, it varied from 0.83 m · s$^{-1}$ for C1 up to 3.32 m · s$^{-1}$ for C4. Similar results were observed at 3 T. The measurements for different phantoms did not overlap at $f_{opt}$ while they did at $f_{conv}$.

Measurement conditioning

When downsampling datasets at $f_{conv}$ in order to achieve optimal $s$, MRE shear velocities increased and the associated variability reduced towards the values found at $f_{opt}$ (Table 2). The Wilcoxon tests revealed statistically significant differences between the four phantom measurements with infinitesimal p-values for any combination of phantom results. After appropriate downsampling, the phantoms could be discriminated (Figure 6 (d)).

### 3.4. MRE and SWE comparison

As expected in the four phantoms, both ultrasound Bmode images and elasticity maps were homogeneous (Figure 7). Measurement variabilities were small and below 6 % except for C1 where it reached 10 % with the SL10-2 probe. Shear velocities obtained with SWE were similar regardless of the probe used (Table 3). Small differences only appeared for the softest phantoms (C1-C2). The overall mean shear velocities were $0.98 \pm 0.04$, $1.63 \pm 0.05$, $2.46 \pm 0.03$ and $3.46 \pm 0.06$ m · s$^{-1}$ for C1, C2, C3 and C4 respectively. The MRE mean shear velocities, averaged between 1.5 T and 3 T ($0.84 \pm 0.05$, $1.51 \pm 0.03$, $2.29 \pm 0.06$ and $3.31 \pm 0.05$ m · s$^{-1}$ for C1, C2, C3 and C4 respectively), compared rather well with the SWE mean shear velocities. Relative discrepancies were below 8 % except for C1, where it reached 18 %. This trend propagated to the shear elasticities with twice as large relative discrepancies, ranging from 10 % to 39 % (Table 3).

The estimated Young's moduli ranged from 2.1 to 33.2 kPa for MRE and from 2.8 to 35.3 kPa for SWE, which were well below the CIRS calibrated values (3.5 to 44.8 kPa), which differed from SWE by 17 % to 53 % and from MRE by 34 % to 80 %.

4. Discussion

MRE was repeated and reproduced in two sites at different magnetic field strengths on four phantoms mechanically excited from 40 Hz to 320 Hz. Although the amplitude of the mechanical waves and the SNRs differed between the platforms, the shear velocities and elasticities matched within the measurement uncertainty for each excitation frequency. Both mean values and standard deviations agreed at 1.5 T and 3 T (Table 1 and Table 2). Despite an average factor three in data quality between 1.5 T and 3 T, the voxel-wise velocities followed the same distribution with respect to $Q$ (Figures S1-S2). Such robustness effectively held when the data were well conditioned but failed otherwise. When $6 \lesssim s \lesssim 9$, both measurement accuracy and precision were preserved at 1.5 T and 3 T with dissimilar $Q$ whereas, when $s \lesssim 6$ or $s \gtrsim 9$, they were degraded. These results fully corroborate the simulations carried out in a purely elastic, isotropic, homogeneous medium.[7] Here, the shear viscosity could be neglected and the four phantoms could be considered as purely elastic in the range of frequencies explored by both MRE and SWE. Indeed, the phase velocities measured at a single excitation frequency with MRE were similar to the group velocities measured within a broadband excitation with SWE. Moreover, the SWE measurements were rather independent of the ultrasound probe central frequencies (3.5, 6.0 and 7.5 MHz) to which the excitation bandwidths are related. We may thus fairly assume that the phantoms are not dispersive as formerly stated on similar Zerdine® solid elastic hydrogels.[18] Therefore, the dispersion law should be a constant and not a monotonically increasing function of frequency as reported here for C2 between 40 and 320 Hz (Figure 4 (a) and (b)). This dispersion is an overall measurement artefact as the increasing trend vanishes when optimal spatial sampling is performed before data processing (Figure 4 (c)). Hence, the increase of shear velocity from 1.44 to 2.53 m·s$^{-1}$ does not reflect the mechanical response of the medium only but instead the bias added along the reconstruction, which depends on the spatial sampling.

In our study, this bias was negative when $s \gtrsim 9$ – here when $f \lesssim 90$ Hz – and positive when $s \lesssim 6$ – here when $f \gtrsim 207$ Hz. Furthermore, this reconstruction bias came with an escalating measurement variability. Even with high $Q$, the measurement precision was progressively degraded as we moved away from the optimal $s$ conditions (Table 1). This is exemplified at both fields where the lowest standard deviations that the measurements exhibited were found at 175 Hz (1.3 % at 1.5 T and 2.6 % at 3 T respectively), compared to much higher standard deviations at 40 Hz (32 % and 27 %) or even at 90 Hz (3.4 % and 6.8 %) although $Q$s were smaller at 175 Hz than at 40 Hz and 90 Hz.

The comparison at different fields might have been weakened by experimental reproducibility issues. The SNR gain of 1.4 instead of 2 as expected at 3 T can be explained by magnetic defects, most probably air bubbles, revealed in the phantoms through a slight granularity of the magnitude images or SNR maps (Figure 3 and Figure 5). The wave amplitude was lower than expected. Although the general arrangement of the two MRI facilities was different, we made sure that the waveguides were of the same length to obtain identical resonant modes. Excitation frequencies were also identical on both platforms. Moreover, we verified that the applied pressures, as optically recorded at the surface of the phantoms, were the same for each frequency on both sites. Yet, the displacement fields measured with MRE did not exhibit the same amplitudes at 1.5 T and 3 T. We assume that the 1 mm diameter acoustic adapter was not properly sealed at 3 T and pressure leaked. Therefore, data quality was degraded in average by a factor 3 at 3 T.

The higher amplitudes at 1.5 T revealed underlying interference patterns that resulted from wave reflections on the cylindrical wall boundary. Spherical spreading did not suffice to attenuate the waves before they bounced onto the wall and too little (if any) attenuation came from the viscosity of the phantoms. This corroborates the purely elastic behavior of the phantoms.

These wave interference patterns have been well described by Okamoto *et al.* in a similar setup with the sum of Bessel functions of the first and second kinds.[22] However, in our study, patterns were not propagated by the reconstruction and they were barely seen on the inferred maps of shear velocity and elasticity (Figure 3 and Figure 5). These results underscore the robustness of AIDE with respect to boundary conditions, multiple reflections, and interferences provided that the waves do not get fully annihilated.[9] The robustness is confirmed by the agreement between MRE and SWE, for which the extraction of the shear velocity is not subjected to any boundary condition. It is obtained here when MRE is performed in optimal conditions and because phase and group velocities match as the phantoms are largely elastic and negligibly viscous.[19]

The MRE shear velocity mean values, estimated in the optimal $s$ domain, stand below the SWE mean values, within 8 %, for C2-C4, at the limit of the added measurement uncertainties. The greater underestimation we report for C1, 18 %, might partly originate from the sur-optimal conditioning ($s = 11.1$) as the optimal conditioning had merely been estimated for C1 (as for C3 and C4) *a priori* from the manufacturer's specifications and not experimentally determined as for C2. Besides, the general negative bias might mainly originate not from MRE underestimation but from SWE overestimation. The SWE overestimation has already been raised by Oudry et al.[23] and by Urban et al.[24] with relative differences of up to 22 % at 400 Hz. Our results question the calibrated Young's moduli provided by the phantom manufacturer, which match neither SWE nor MRE inferred values. Indeed, the manufacturer Young's moduli do increase with the phantom stiffness but largely overestimate SWE Young's moduli by 15 % to 35 % and MRE Young's moduli by 25 % to 44 %.

In the work by Oudry et al.[23], MRE and ultrasound transient elastography (TE) measurements were averaged over multi-frequency acquisitions on four styrene-ethylene/butylene-styrene phantoms between 60 Hz and 220 Hz. The phantoms were considered more elastic than viscous and shear elasticities were averaged over the frequency range for MRE-TE comparison. Yet, we can estimate that, over the explored ranges, $s$ roughly spans from 17.2 down to 6.5 for the softer phantom and from 17.7 down to 12.9 for the stiffer phantom. Namely, while sweeping the frequency spectrum, the optimal conditions are not always fulfilled and MRE shear elasticities are biased negatively and positively with respect to the effective $s$ at the applied excitation frequency. No special spectral trend emerges for the stiffer phantom as all shear elasticity values are recorded outside the optimal domain with expectedly measurement uncertainties overwhelming the trend. Yet, noticeable increasing trends come as a rather clear signature for the other three phantoms.

Bigot et al. carried out a thorough comparative study on agarose phantoms with inclusions of two types of cerebral fibrils and bovine serum albumin (BSA).[17] MRE was performed at multiple frequencies between 400 Hz and 1200 Hz. Shear velocities increased over the frequency span from $(2.01 \pm 0.77)$ m·s⁻¹ to $(2.59 \pm 0.42)$ m·s⁻¹ in average in fibrils and from $(2.28 \pm 0.69)$ m·s⁻¹ to $(2.57 \pm 0.48)$ m·s⁻¹ in BSA. The voxel size of the acquisitions was 0.391 mm and the spatial sampling factors below 800 Hz ranged out of the optimal domain ($s \gtrsim 10.3$) as substantiated by nearly twice larger associated measurement standard deviations. Therefore, we do speculate that at least part, if not all, of the reported dispersive behaviors of fibrils and BSA could be explained by the positive measurement bias expected in these conditions.

In this framework, we could interestingly review a pioneering work that was carefully performed by Green et al. with roughly $s \simeq 7$ at a single frequency in a gelatin phantom with four mechanically different regions.[15] Nevertheless, the reported shear elasticities were systematically lower with MRE at 200 Hz, $G'_{\mathrm{MRE}} = \{6.6, 12.0, 16.2, 23.0\}$ kPa, than with rheometry at 50 Hz, $G'_{\mathrm{Rheo}} = \{6.7, 14.2, 24.2, 33.2\}$ kPa. The negative MRE measurement bias remained effectively small, within 15 %, as long as $s$ remained within the optimal domain, which held for the two softer regions of the phantom ($s \lesssim 9$). However, for the two stiffer regions, $s$ was beyond 10 and the measurement bias was above 30 %. Our interpretation is also confirmed by the increasing measurement deviations for stiffer regions – from 2 % to 5 % and 10 %.

The data conditioning pitfall is exemplified in this work by the comparison between MRE at $f_{\mathrm{conv}} = 60$ Hz (Figure 6 (c)), which is currently established as the reference procedure for diagnosing liver fibrosis,[3,25,26] and the proposed $s$-optimized multi-frequency MRE (Figure 6 (a) or (b)) or multi-sampling MRE (Figure 6 (d)). When data are not well conditioned, namely here at $f_{\mathrm{conv}}$, when $s$ goes farther away from the optimal domain from $s \simeq 11.1$ for C1 up to $s \simeq 43.9$ for C4 (Table 2), the shear velocity mean value gets so underestimated, down to 43.2 %, and the measurement variability gets so large, up to fivefold, that the rather far apart fibrosis severity mimicked by the C3 and C4 phantoms cannot be mechanically discriminated (Figure 6 (c)). When data are well conditioned, either prospectively by adjusting the excitation frequencies and the voxel size with respect to the expected shear wavelengths or retrospectively by resampling the recorded MRI data with respect to the estimated shear wavelengths, then MRE measurement uncertainty is minimized and shear velocity values are as much accurate and precise as they can be with the available data quality. In these conditions only, correct quantitative MRE can be achieved at the voxel level.

The prospective adjustment of a single excitation frequency may not be clinically applicable, since the tissue stiffness is not known a priori and is in contrary the unknown under consideration. Yet it is possible to set a broad spectrum of optimal conditions that would cover the expected mechanical range of an organ or a disease by implementing multi-frequency acquisitions.[11,27–31] For each frequency, a different velocity map would be estimated with the same voxel size. Mechanically-homogenous regions could then be selected from the velocity map corresponding to the central frequency map using e.g., k-means clustering. Then, for each frequency, the standard deviation of the shear velocity could be computed in the different segmented regions and an optimal shear velocity map could be composed by individually choosing for each region, the velocity obtained at the frequency that provides the lowest associated standard deviation. Providing that the multi-frequency acquisition covers a wide enough range of frequencies, the resulting shear velocity map would achieve piecewise optimal conditions within the assumptions of dominantly elastic, locally homogenous, and isotropic medium.

Alternatively, with standard single frequency MRE, multi-resampling of the displacement field maps could be implemented before processing the mechanical parametric maps. The resulting maps could then be resampled back to the original matrix and voxel sizes and mechanically-homogenous regions could be selected across the set of parametric maps. Along the same minimization process described above for multi-frequency MRE, a composite shear velocity map could eventually be established at the acquisition frequency with piecewise optimal conditions within the assumptions of locally homogenous and isotropic medium. To avoid any loss of spatial mechanical differentiation when small regions are targeted, highly spatially resolved displacement fields should be acquired with accordingly high excitation frequency to achieve optimal MRE in the expectedly stiffest regions such that only piecewise interpolation will remain to be performed in the other softer regions.

The multi-frequency prospective approach could be refined by retrospective multi-resampling to cope with an acquisition with a limited range of frequencies or a medium with a broad velocity distribution. The frequency range and the voxel size should be taken according to the SNR and the size of the regions to be characterized. Indeed, the retrospective approach may hinder the effective spatial resolution of MRE when data downsampling is required. In our study, we even reached the bottom line at 40 Hz as we were left with only a single reconstructed slice after reconditioning the data set from 1.25 mm to 5.70 mm. In this case, measurement accuracy and precision come at the expense of spatial resolution and the tradeoff should be thoroughly studied on heterogeneous media before deciding upon optimal parameters to implement for MRE acquisition.

## 5. Conclusion

Mechanical properties assessed with MRE may provide critical insights into the tissue pathophysiological state. The compelling dynamic and dispersive characterization of biological tissues with MRE does not come however without any pitfall. Preferably, the data should be of high quality and the displacement fields induced in the targeted homogeneous tissue at a single frequency should be sampled with the optimal number of voxels per expected wavelength ($6 \lesssim s \lesssim 9$). Yet, acquired data are usually of acceptable but not exceptional quality and the tissue is generally heterogeneous requiring multiple optimal $s$ domains throughout the tissue that cannot be achieved altogether. Absolute quantification can only be completed when optimal conditions are fulfilled either prospectively by adequate multi-frequency excitation or retrospectively by data multi-resampling. Once achieved, wherever stands the quality of available data, velocity standard deviations in mechanically-homogeneous sub-regions will serve as markers of accuracy: the lower the deviation, the better the accuracy. MRE measurement accuracy and precision will then be optimal such that intra-subject or inter-subject regional or temporal tissue mechanical variations can be quantified and discriminated as shown here at the voxel level on phantoms mimicking liver fibrosis.

## 6. Acknowledgements

The MRE experiments were performed on the 1.5 T MRI platform of CEA/SHFJ and the 3 T MRI platform of Beaujon Hospital affiliated to the France Life Imaging network (grant ANR-11-INBS-0006).

The authors are thankful to Jean-Luc Gennisson for fruitful discussions on SWE.

Elasticity quantitation and tissue discrimination with MRE

Table 1: MRE acquisition parameters and MRE mechanical outcomes for the CIRS liver fibrosis phantoms C2 at 1.5 T and 3 T. The voxel size is 1.25 mm for all experiments. $B_0$: magnetic field strength, $TR/TE$: repetition and echo times, $a$: voxel size, $f$: excitation frequency, $s$: sampling factor or number of voxels per wavelength, SNR: signal-to-noise ratio, $A$: displacement field amplitude, $Q$: quality factor, $V_s$: shear velocity, and $G'$: shear elasticity.

| CIRS | $B_0$ [T] | TR/TE [ms] | $a$ [mm] | $f$ [Hz] | $\langle s \rangle$ | $\langle SNR \rangle$ | $\langle A \rangle$ [μm] | $\langle Q \rangle$ | $\langle V_s \rangle$ [m·s⁻¹] | $\langle G_d \rangle$ [kPa] |
|---|---|---|---|---|---|---|---|---|---|---|
| C2 | 1.5 | 1800/62 | 1.25 | 40 | 30.2 | 29 | 12.36 ± 3.74 | 117 ± 44 | 1.47±0.47 | 2.38±3.30 |
| C2 | 3.0 | 1800/62 | 1.25 | 40 | 30.2 | 45 | 4.50 ± 1.32 | 63 ± 31 | 1.48±0.40 | 2.34±1.60 |
| C2 | 1.5 | 1600/58 | 1.25 | 60 | 20.1 | 29 | 15.03 ± 4.78 | 137 ± 51 | 1.44±0.12 | 2.10±0.35 |
| C2 | 3.0 | 1333/39 | 1.25 | 60 | 20.1 | 40 | 1.93 ± 0.66 | 30 ± 11 | 1.44±0.33 | 2.20±1.50 |
| C2 | 1.5 | 1600/50 | 1.25 | 90 | 13.4 | 30 | 13.99 ± 6.06 | 274 ± 124 | 1.46±0.05 | 2.15±0.16 |
| C2 | 3.0 | 1200/42 | 1.25 | 90 | 13.4 | 45 | 3.27 ± 0.73 | 43 ± 17 | 1.46±0.10 | 2.14±0.30 |
| C2 | 1.5 | 1477/46 | 1.25 | 130 | 9.3 | 30 | 3.57 ± 1.23 | 67 ± 26 | 1.48±0.04 | 2.19±0.12 |
| C2 | 3.0 | 923/27 | 1.25 | 130 | 9.3 | 34 | 3.36 ± 0.86 | 37 ± 17 | 1.49±0.05 | 2.23±0.16 |
| C2 | 1.5 | 1477/46 | 1.25 | 130 | 9.3 | 30 | 3.27 ± 1.83 | 68 ± 27 | 1.48±0.04 | 2.19±0.12 |
| C2 | 3.0 | 923/27 | 1.25 | 130 | 9.3 | 34 | 3.04 ± 1.34 | 40 ± 17 | 1.50±0.05 | 2.25±0.16 |
| C2 | 1.5 | 1370/46 | 1.25 | 175 | 6.9 | 27 | 4.46 ± 2.18 | 104 ± 42 | 1.51±0.02 | 2.27±0.07 |
| C2 | 3.0 | 1096/37 | 1.25 | 175 | 6.9 | 41 | 0.46 ± 0.15 | 17 ± 7 | 1.51±0.04 | 2.30±0.13 |
| C2 | 1.5 | 1158/36 | 1.25 | 207 | 5.8 | 24 | 2.03 ± 1.36 | 34 ± 17 | 1.53±0.04 | 2.33±0.12 |
| C2 | 3.0 | 1042/36 | 1.25 | 207 | 5.8 | 39 | 0.53 ± 0.19 | 15 ± 6 | 1.56±0.05 | 2.42±0.16 |
| C2 | 1.5 | 1049/34 | 1.25 | 320 | 3.8 | 22 | 1.63 ± 0.55 | 5 ± 3 | 2.53±0.50 | 6.64±3.80 |
| C2 | 3.0 | 1049/34 | 1.25 | 320 | 3.8 | 39 | 3.15 ± 0.93 | 9 ± 10 | 2.13±0.40 | 4.40±2.50 |

Table 2: MRE acquisition parameters and MRE mechanical outcomes for the four CIRS liver fibrosis phantoms C1-C4 at 1.5 T (light blue) and 3 T (dark blue) at optimal frequencies $f_{opt} = \{60, 175, 207, 320\}$ Hz as well as at 1.5 T, 60 Hz without (white) and with resampling (light green). $B_0$: magnetic field strength, $f$: excitation frequency, $TR/TE$: repetition and echo times, $a$: voxel size, $s$: sampling factor or number of voxels per wavelength, SNR: signal-to-noise ratio, $A$: displacement field amplitude, $Q$: quality factor, $V_s$: shear velocity, $G'$: shear elasticity, and $E$: Young's modulus

| CIRS | $B_0$ [T] | TR/TE [ms] | $a$ [mm] | $f$ [Hz] | $\langle s \rangle$ | $\langle SNR \rangle$ | $\langle A \rangle$ [μm] | $\langle Q \rangle$ | $\langle V_s \rangle$ [m·s$^{-1}$] | $\langle G_d \rangle$ [kPa] | $\langle E \rangle$ [kPa] |
|---|---|---|---|---|---|---|---|---|---|---|---|
| C1 | 1.5 | 1600/58 | 1.25 | 60  | 11.1 | 30 | 10.13 ± 4.35 | 114 ± 40  | 0.83 ± 0.06 | 0.69 ± 0.12 | 2.07 |
| C1 | 3.0 | 1200/42 | 1.25 | 60  | 11.1 | 38 | 1.25 ± 0.59  | 21 ± 8    | 0.84 ± 0.07 | 0.71 ± 0.12 | 2.13 |
| C2 | 1.5 | 1370/46 | 1.25 | 175 | 6.9  | 27 | 4.49 ± 2.17  | 105 ± 43  | 1.51 ± 0.02 | 2.27 ± 0.07 | 6.81 |
| C2 | 3.0 | 1096/37 | 1.25 | 175 | 6.9  | 41 | 0.46 ± 1.15  | 17 ± 7    | 1.51 ± 0.04 | 2.30 ± 0.13 | 6.90 |
| C2 | 1.5 | 1600/58 | 1.25 | 60  | 20.1 | 26 | 15.03 ± 4.78 | 137 ± 51  | 1.44 ± 0.12 | 2.10 ± 0.35 | 6.30 |
| C2 | 1.5 | 1096/37 | 3.63 | 60  | 6.9  | 31 | 0.46 ± 1.15  | 17 ± 7    | 1.44 ± 0.03 | 2.30 ± 0.13 | 6.90 |
| C3 | 1.5 | 1158/36 | 1.25 | 207 | 8.7  | 26 | 2.85 ± 1.10  | 65 ± 26   | 2.27 ± 0.08 | 5.15 ± 0.37 | 15.45 |
| C3 | 3.0 | 1042/36 | 1.25 | 207 | 8.7  | 40 | 0.92 ± 0.34  | 32 ± 16   | 2.30 ± 0.08 | 5.32 ± 0.39 | 15.96 |
| C3 | 1.5 | 1200/42 | 1.25 | 60  | 29.9 | 29 | 8.69 ± 4.13  | 59 ± 24   | 1.76 ± 0.50 | 3.34 ± 2.13 | 10.02 |
| C3 | 1.5 | 1200/42 | 4.29 | 60  | 8.7  | 29 | 7.24 ± 2.79  | 173 ± 70  | 2.20 ± 0.07 | 4.82 ± 0.31 | 14.46 |
| C4 | 1.5 | 1049/34 | 1.25 | 320 | 8.3  | 24 | 6.49 ± 2.55  | 132 ± 60  | 3.32 ± 0.09 | 11.03 ± 0.59 | 33.30 |
| C4 | 3.0 | 1049/34 | 1.25 | 320 | 8.3  | 41 | 5.37 ± 2.09  | 179 ± 106 | 3.30 ± 0.06 | 10.09 ± 0.39 | 30.27 |
| C4 | 1.5 | 1200/42 | 1.25 | 60  | 44.4 | 29 | 12.59 ± 8.00 | 61 ± 37   | 1.88 ± 0.77 | 4.13 ± 4.50 | 12.39 |
| C4 | 1.5 | 1200/42 | 5.70 | 60  | 9.7  | 29 | 17.38 ± 6.09 | 348 ± 95  | 2.96 ± 0.19 | 8.81 ± 1.11 | 26.43 |

Table 3: Shear velocity mean values and standard deviations for the four CIRS liver fibrosis phantoms C1-C4 reported with MRE at 1.5 T and 3 T at conventional excitation frequency $f_{\text{conv}} = 60$ Hz, without and with data resampling, and at optimal excitation frequencies $f_{\text{opt}} = \{60, 175, 207, 320\}$ Hz; and by SWE with three ultrasound probes, XC6-1, SL10-2, and SL15-4 at 3.5, 6, and 7.5 MHz respectively.

| CIRS | $\langle V_s \rangle_{\text{MRE}}$ [m·s⁻¹] | | | | $\langle V_s \rangle_{\text{SWE}}$ [m·s⁻¹] | | |
|---|---|---|---|---|---|---|---|
| | $B_0 = 1.5$ T<br>$a = 1.25$ mm<br>$f = f_{\text{conv}}$ | $B_0 = 1.5$ T<br>$a = a_{\text{opt}}$<br>$f = f_{\text{conv}}$ | $B_0 = 1.5$ T<br>$a = 1.25$ mm<br>$f = f_{\text{opt}}$ | $B_0 = 3$ T<br>$a = 1.25$ mm<br>$f = f_{\text{opt}}$ | Probe<br>XC6-1 MHz | Probe<br>SL10-2 MHz | Probe<br>SL15-4 MHz |
| C1 | 0.83 ± 0.06 | 0.83 ± 0.06 | 0.83 ± 0.06 | 0.84 ± 0.07 | 0.96 ± 0.03 | 1.01 ± 0.11 | 0.97 ± 0.05 |
| C2 | 1.44 ± 0.12 | 1.44 ± 0.03 | 1.51 ± 0.02 | 1.51 ± 0.04 | 1.60 ± 0.02 | 1.66 ± 0.10 | 1.61 ± 0.10 |
| C3 | 1.76 ± 0.50 | 2.20 ± 0.07 | 2.27 ± 0.08 | 2.30 ± 0.08 | 2.43 ± 0.04 | 2.47 ± 0.06 | 2.48 ± 0.08 |
| C4 | 1.88 ± 0.77 | 2.96 ± 0.19 | 3.32 ± 0.09 | 3.30 ± 0.06 | 3.51 ± 0.11 | 3.43 ± 0.12 | 3.43 ± 0.10 |

Table 4: MRE acquisition parameters and MRE mechanical observables for the CIRS liver fibrosis phantom C2 at 1.5 T after resampling the raw displacement fields to reach the optimal $s$ domain with voxel size $a = a_{\text{opt}}$ instead of $a = 1.25$ mm. $B_0$: magnetic field strength, $TR/TE$: repetition and echo times, $a_{\text{opt}}$: voxel size, $f$: excitation frequency, $s$: sampling factor or number of voxels per wavelength, SNR: signal-to-noise ratio, $A$: displacement field amplitude, $Q$: quality factor, $V_s$: shear velocity, and $G'$: shear elasticity.

| CIRS | $B_0$ [T] | TR/TE [ms] | $a_{\text{opt}}$ [mm] | $f$ [Hz] | $\langle s \rangle$ | $\langle SNR \rangle$ | $\langle A \rangle$ [µm] | $\langle Q \rangle$ | $\langle V_s \rangle$ [m·s$^{-1}$] | $\langle G' \rangle$ [kPa] |
|---|---|---|---|---|---|---|---|---|---|---|
| C2 | 1.5 | 1800/62 | 5.00 | 40 | 7.0 | 31 | 8.92 ± 2.47 | 450 ± 150 | 1.40±0.02 | 1.97±0.06 |
| C2 | 1.5 | 1600/58 | 3.33 | 60 | 6.9 | 30 | 11.44 ± 3.23 | 289 ± 98 | 1.40±0.03 | 1.88±0.09 |
| C2 | 1.5 | 1600/50 | 2.31 | 90 | 6.95 | 31 | 9.32 ± 3.83 | 323 ± 132 | 1.44±0.03 | 2.08±0.09 |
| C2 | 1.5 | 1477/46 | 1.67 | 130 | 6.9 | 31 | 2.28 ± 0.79 | 56 ± 23 | 1.49±0.02 | 2.21±0.08 |
| C2 | 1.5 | 1370/46 | 1.25 | 175 | 6.9 | 24 | 2.95 ± 1.42 | 63 ± 25 | 1.51±0.02 | 2.27±0.07 |
| C2 | 1.5 | 1158/36 | 1.03 | 207 | 7.2 | 23 | 1.48 ± 0.94 | 26 ± 13 | 1.53±0.04 | 2.35±0.11 |
| C2 | 1.5 | 1049/34 | 0.68 | 320 | 7.3 | 21 | 1.21 ± 0.32 | 11 ± 4 | 1.59±0.12 | 2.55±0.40 |

Elasticity quantitation and tissue discrimination with MRE

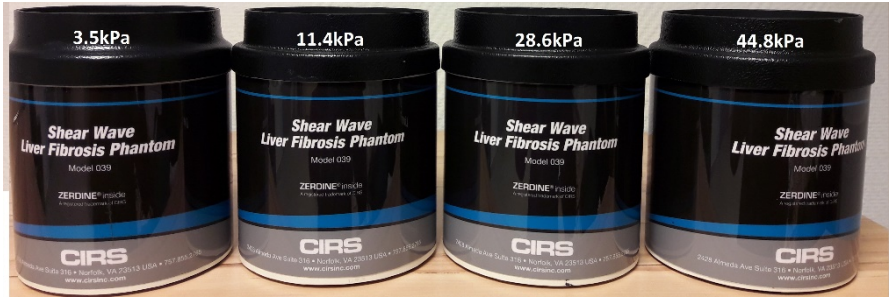

Figure 1: Set of CIRS test phantoms {$C1, C2, C3, C4$} calibrated for liver fibrosis at 3.5, 11.4, 28.6 and 44.8 kPa

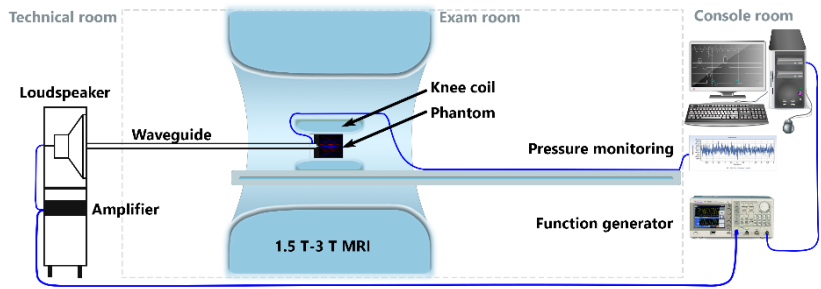

Figure 2: Schematic of the MRE bench setup at 1.5 T and 3 T. Remotely generated and amplified pressure waves are guided in the center of the MRI magnet bore to the phantom placed in a knee coil. Pressure level is monitored via an optical fiber sensor. Pressure waves are synchronized to the MRI acquisition sequence.

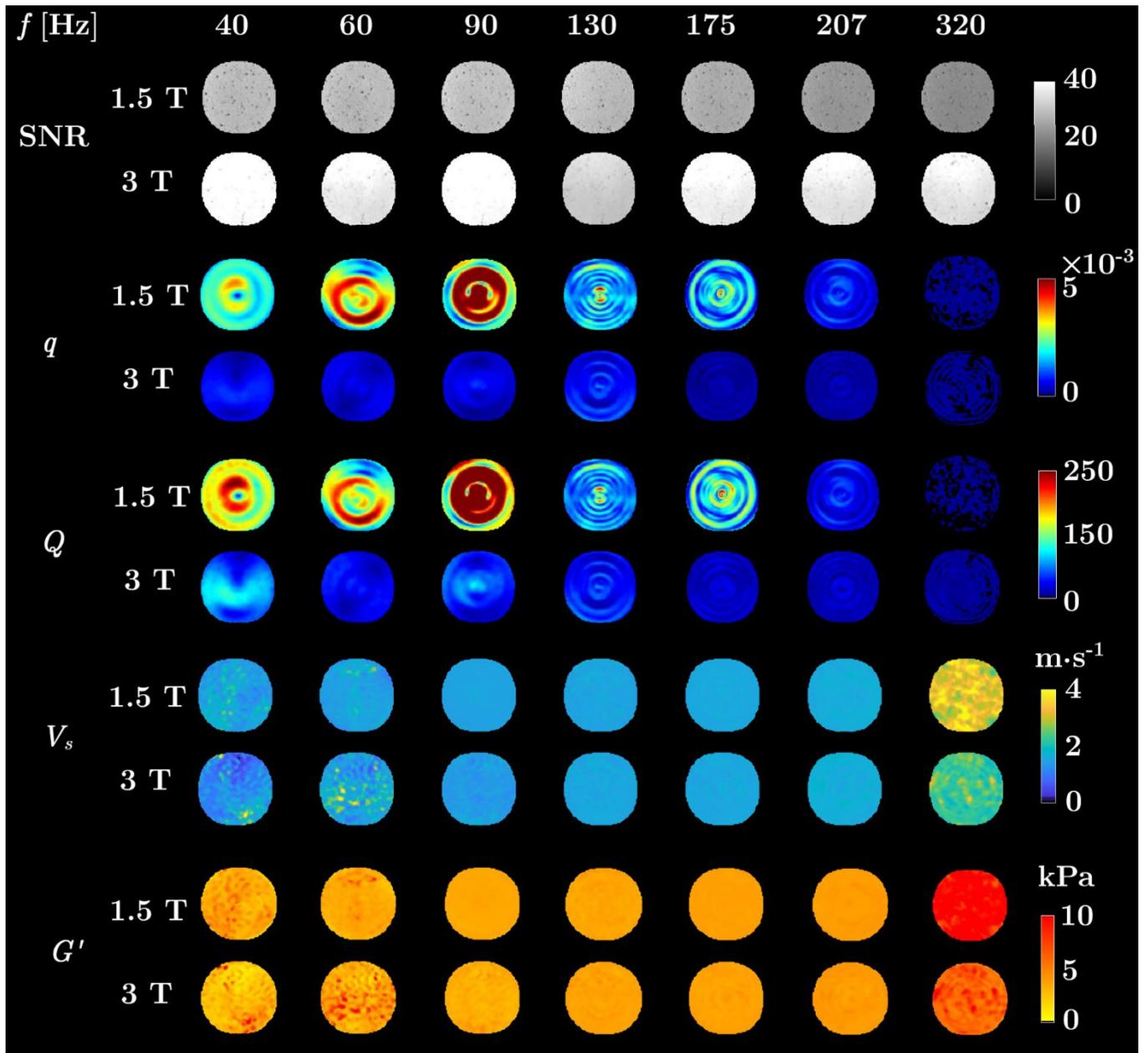

Figure 3: SNR, amplitude of the curl of the displacement field $q$, data quality $Q$, shear velocity $V_s$, and shear elasticity $G'$ maps for the CIRS liver fibrosis phantom C2 with MRE at excitation frequencies $f =$ {40,60,90,130,175,207,320} Hz at magnetic field strength of 1.5 T and 3 T. With consistent SNR but inhomogeneous $q$ and $Q$ maps, MRE is well conditioned and provides homogeneous $V_s$ and $G'$ maps in the 90 Hz-207 Hz frequency range.

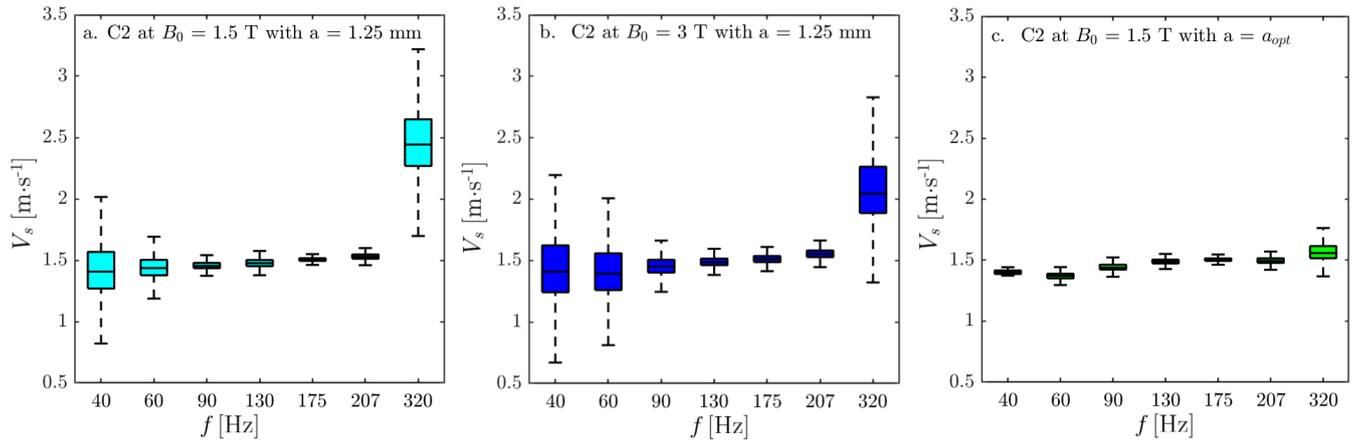

Figure 4: MRE shear velocity $V_s$ as a function of the excitation frequency, $f$, in the CIRS liver fibrosis phantom C2 at 1.5 T (light blue), 3 T (blue) and after data resampling at 1.5 T (green). At both fields, MRE optimal conditions are reached at $f = 175$ Hz where measurement uncertainty is minimized (a and b). At other excitation frequencies, MRE accuracy and precision are restored by data resampling (c).

Elasticity quantitation and tissue discrimination with MRE

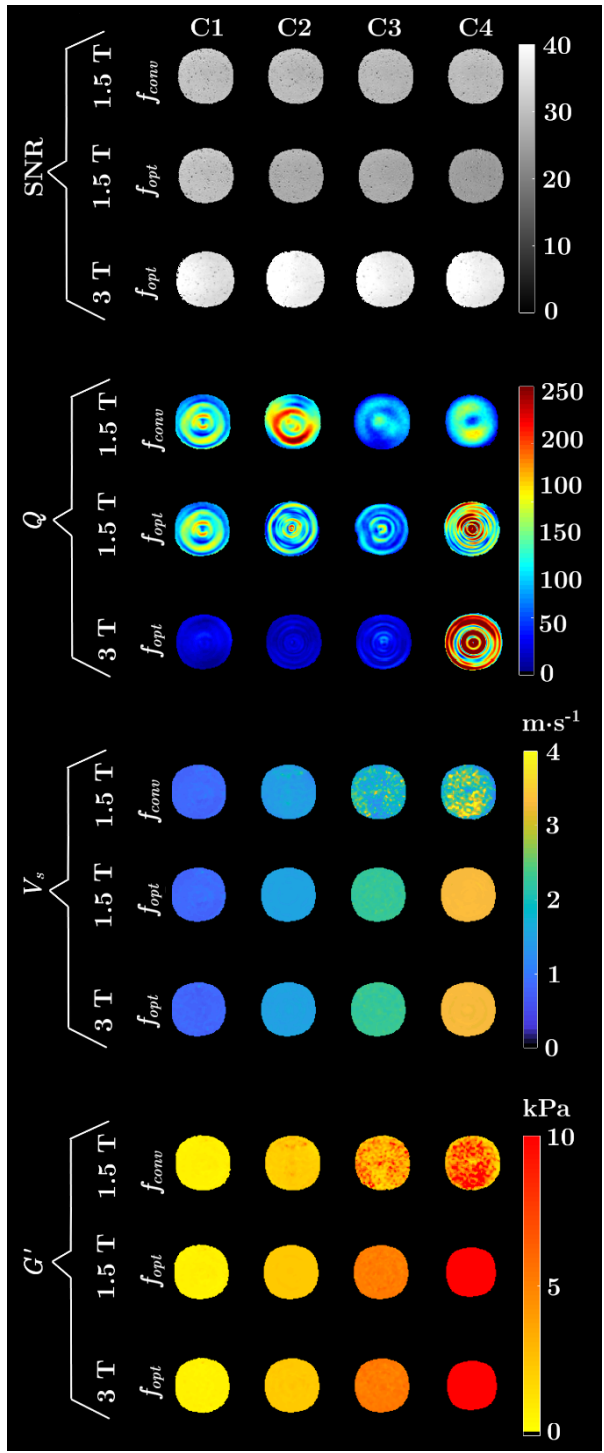

Figure 5: SNR, data quality $Q$, shear velocity $V_s$, and shear elasticity $G'$ maps for the four CIRS liver fibrosis phantoms C1-C4 with MRE at conventional excitation frequency $f_{\text{conv}} = 60$ Hz (for reference, the manufacturer specifications were 1.7, 3.8, 9.5, and 14.9 kPa for shear elastic modulus and 1.1, 1.9, 3.0, and 3.8 m · s$^{-1}$ for shear velocities). Despite inhomogeneous $Q$ maps, MRE is well conditioned and provides homogeneous $V_s$ and $G'$ maps only for C1. It is degraded when departing from these optimal $s$ conditions as the elasticity, hence the wavelength, increases for the three other phantoms C2-C4.

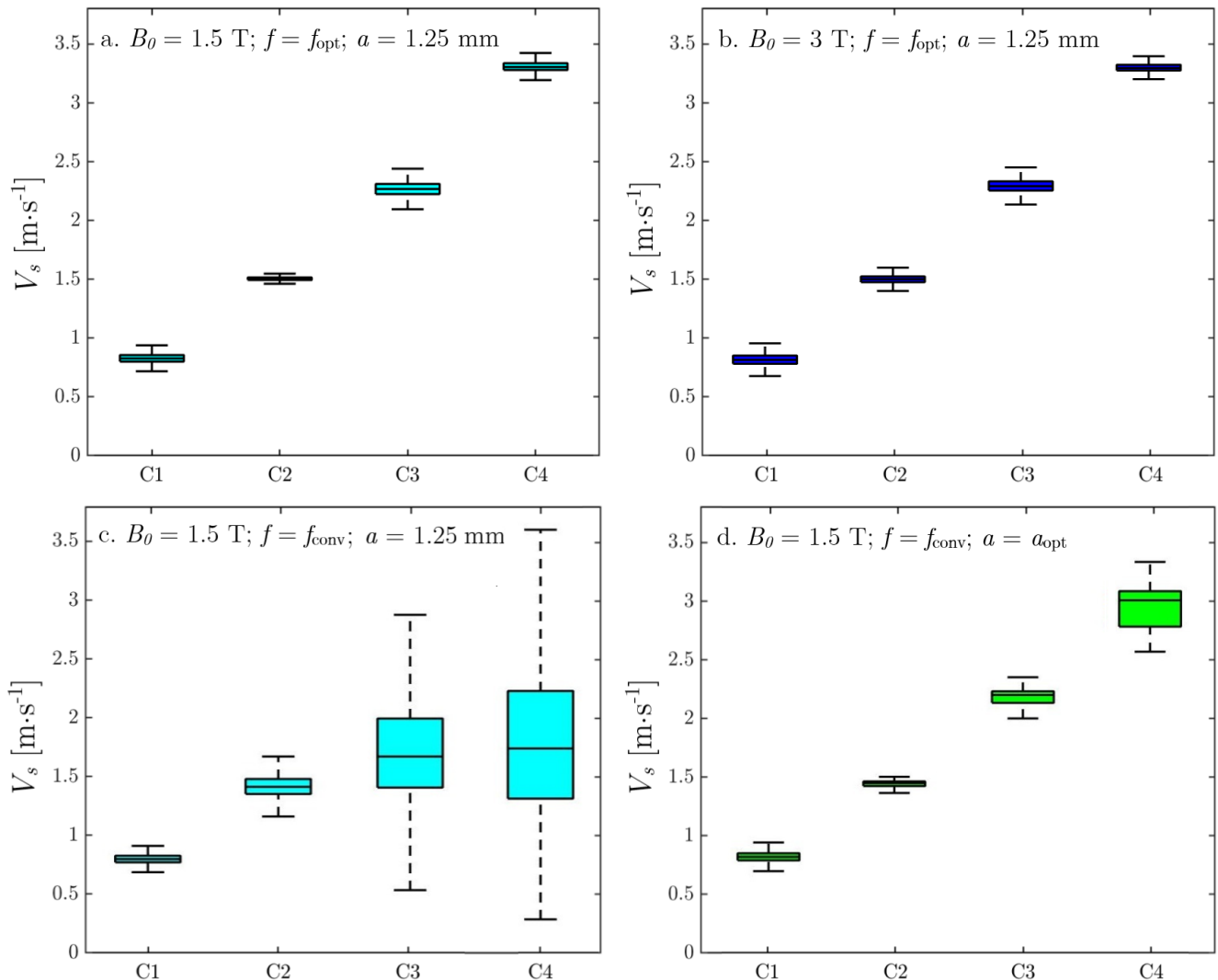

Figure 6: MRE shear velocity $V_s$ in four CIRS liver fibrosis phantoms {C1,C2,C3,C4} at optimal excitation frequencies $f_{\text{opt}} = \{60, 175, 207, 320\}$ Hz at 1.5 T (a) and 3 T (b) and at conventional excitation frequency $f_{\text{conv}} = 60$ Hz without (c) and with data resampling (d). When properly conditioned, either prospectively by multi-frequency acquisition in the optimal $s$ domain (a and b), or retrospectively by multi-resampling of the data to the optimal $s$ domain (d), MRE measurement sensitivity and specificity are improved with regard to conventional MRE at $f_{\text{conv}} = 60$ Hz (c).

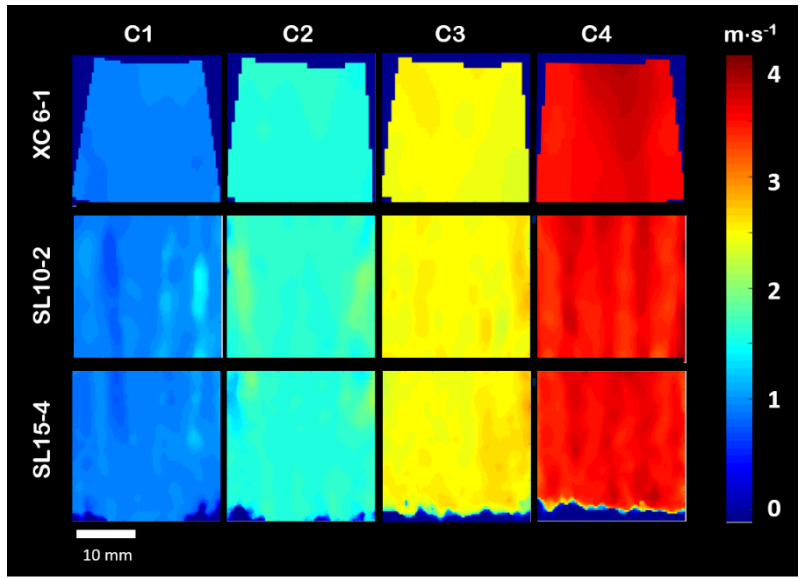

Figure 7: Ultrasound shear velocity maps obtained with three different ultrasound transducers (XC6-1, SL10-2, SL15-4 from top to bottom) for the phantoms C1-C4 (from left to right).

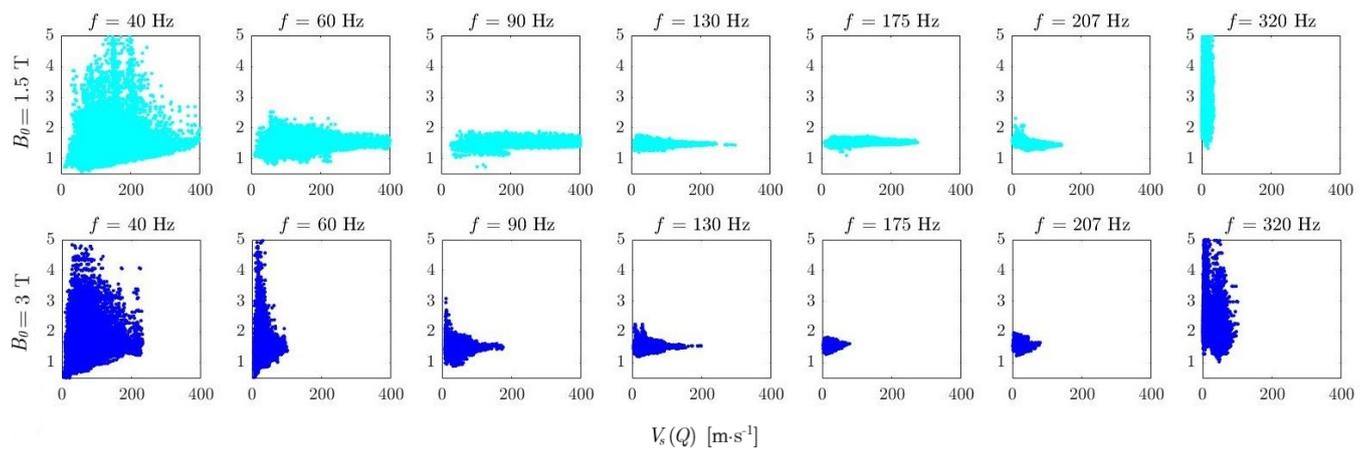

Figure S1: MRE voxel-wise shear velocity $V_s$ as a function of data quality $Q$ in the CIRS liver fibrosis phantom C2 at excitation frequencies $f = \{40, 60, 90, 130, 175, 207, 320\}$ Hz at 1.5 T (top row) and 3 T (bottom row). Velocity values increase with the frequency. Velocity distributions are narrowed in the optimal $s$ domain between 130 Hz and 207 Hz ($6 \lesssim s \lesssim 9$) where the measurements are expected to be the most accurate and precise.

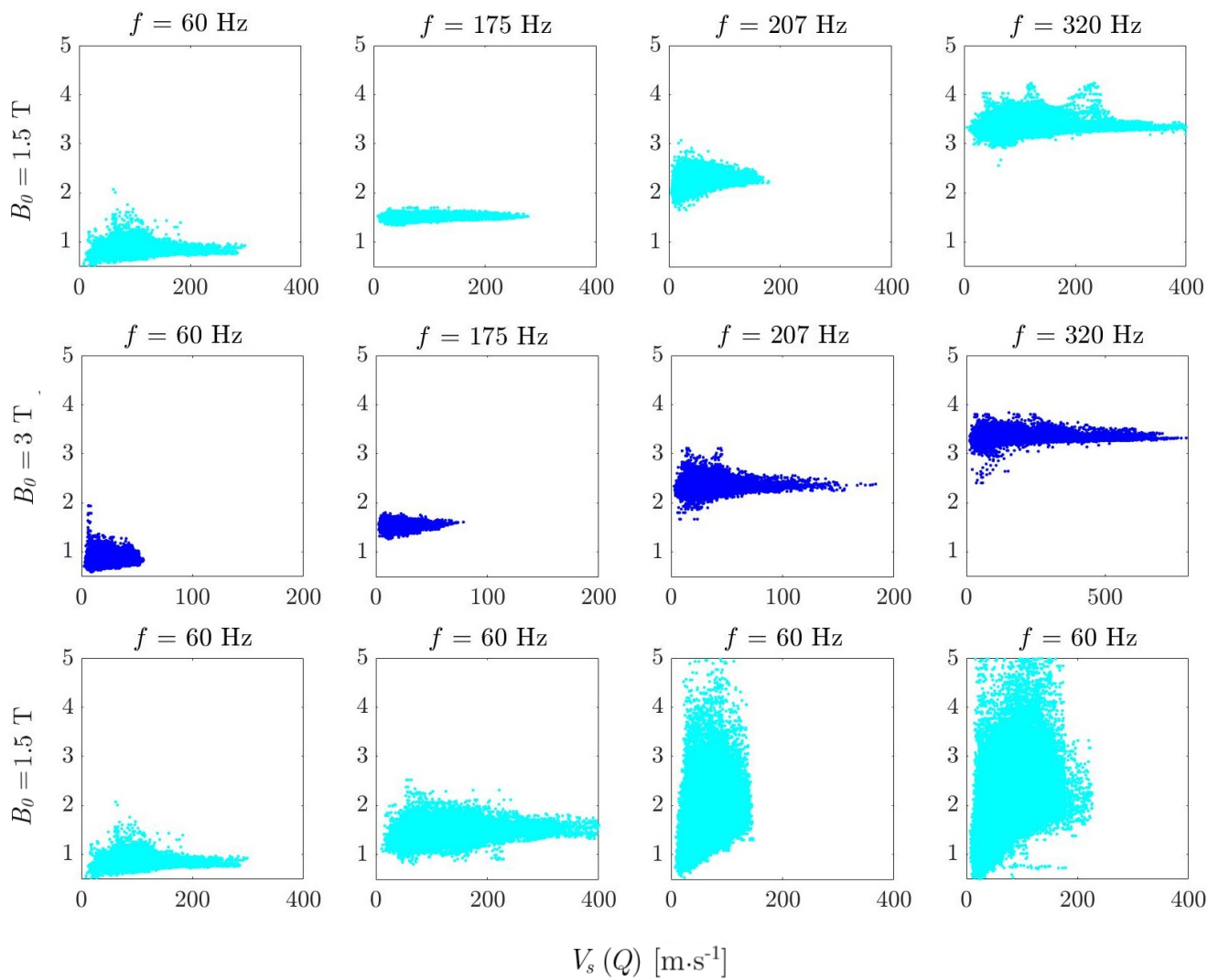

Figure S2: MRE voxel-wise shear velocity $V_s$ as function of $Q$ for the four CIRS liver fibrosis phantoms {C1,C2,C3,C4} at optimal excitation frequencies $f_{\text{opt}} = \{60,175,207,320\}$ Hz for magnetic fields at 1.5 T (upper row) and 3 T (middle row) and at conventional excitation frequency $f_{\text{conv}} = 60$ Hz (bottom row). For both magnetic field strengths, the measurement uncertainties are best when optimal conditions are matched ($6 \lesssim s \lesssim 9$), here at $f_{\text{opt}}$ for each phantom, irrespective of the data quality, $Q$.

Elasticity quantitation and tissue discrimination with MRE

## Multi-frequency MRE for elasticity quantitation and optimal tissue discrimination: a two-platform liver fibrosis mimicking phantom study

Andoh F., Yue J.L., Julea F., Tardieu M., Noûs C., Page G., Garteiser P., Van Beers B.E., Maître X., Pellot-Barakat C.

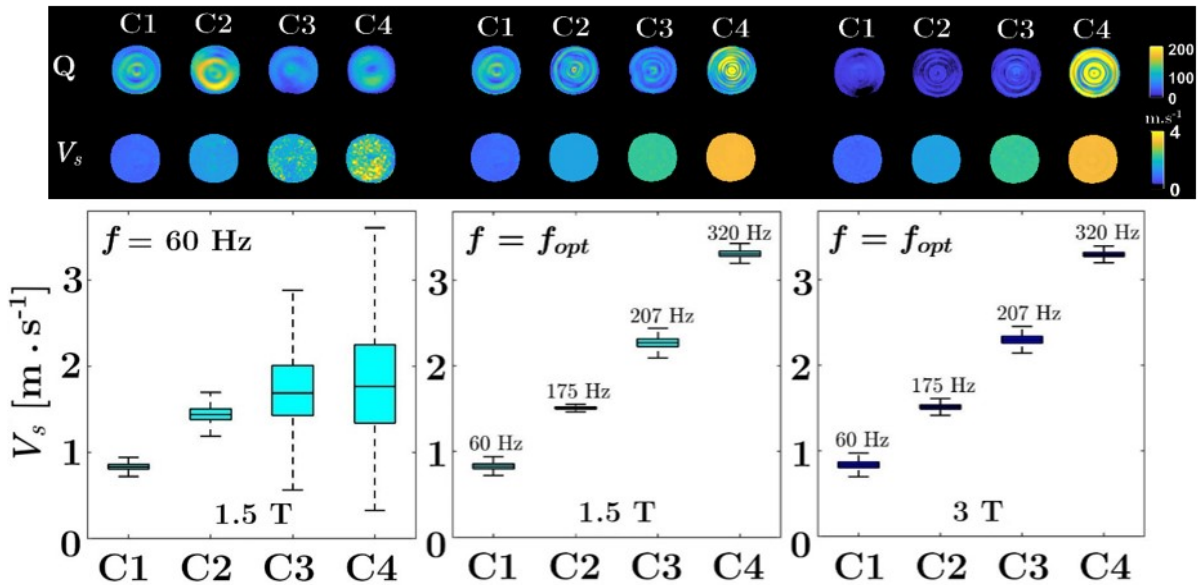

Elastic homogeneous phantoms (C1-C4) of increasing elasticities were mechanically characterized at excitation frequencies $f$ by MRE at 1.5 T and 3 T.

At conventional $f = 60\,\text{Hz}$, estimated shear velocities $V_s$ were highly dispersed leading to overlapping phantom measurements.

At $f = f_{opt}$ (leading to optimal shear wave spatial sampling), measurements were both precise and accurate and phantoms were easily discriminated.

The quality of acquired data $Q$ was higher at 1.5 T than at 3 T, but $V_s$ were consistent for both magnetic fields.